\documentclass[twocolumn,showpacs,preprintnumbers,amssymb,nofootinbib]{revtex4}

\usepackage{graphicx}
\usepackage{bm} 

\begin{document}


\title{Pair creation of higher dimensional black holes on a de Sitter background}

\author{\'Oscar J. C. Dias}
\email{oscar@fisica.ist.utl.pt} \affiliation{ Centro
Multidisciplinar de Astrof\'{\i}sica - CENTRA, Departamento de
F\'{\i}sica, Instituto Superior T\'ecnico, Av. Rovisco Pais 1,
1049-001 Lisbon, Portugal \\ \&\\ CENTRA, Departamento de
F\'{\i}sica, F.C.T., Universidade do Algarve, Campus de Gambelas,
8005-139 Faro, Portugal}
\author{Jos\'e P. S. Lemos}
\email{lemos@kelvin.ist.utl.pt} \affiliation{ Centro
Multidisciplinar de Astrof\'{\i}sica - CENTRA, Departamento de
F\'{\i}sica, Instituto Superior T\'ecnico, Av. Rovisco Pais 1,
1049-001 Lisbon}

\date{\today}
\begin{abstract}
We study in detail the quantum process in which a pair of black
holes is created in a higher $D$-dimensional de Sitter (dS)
background. The energy to materialize and accelerate the pair
comes from the positive cosmological constant. The instantons that
describe the process are obtained from the Tangherlini black hole
solutions. Our pair creation rates reduce to the pair creation
rate for Reissner-Nordstr\"om$-$dS solutions when $D=4$.
Pair creation of black holes in the dS background becomes less
suppressed when the dimension of the spacetime increases. The dS
space is the only background in which we can discuss analytically
the pair creation process of higher dimensional black holes, since
the C-metric and the Ernst solutions, that describe respectively a
pair accelerated by a string and by an electromagnetic field, are
not know yet in a higher dimensional spacetime.
\end{abstract}

\pacs{04.70.Dy, 04.70.-s, 04.20.Gz, 98.80.Jk, 98.80.Cq}


\maketitle

\section{\label{sec:Int}Introduction}

String theories have boosted the interest on higher dimensional
spacetimes. This interest was renewed recently in connection to
the TeV-scale theory \cite{hamed} which suggests that the universe
in which we live may have large extra dimensions. According to
this conjecture, we would live on a four-dimensional sub-manifold,
where the Standard Model inhabits, whereas the gravitational
degrees of freedom propagate throughout all dimensions. Studies on
higher dimensional spacetimes have also raised the possibility
that future accelerators, such as the Large Hadron Collider (LHC)
at CERN produce black holes, and thus detect indirectly
gravitational waves \cite{bhprod}.

Higher dimensional black holes, in particular, are providing a
useful background to test the possible existence of extra
dimensions. It is therefore important to discuss processes that
might lead to the production of black holes in a higher
dimensional background. Among these there is one that deserves a
special attention: the quantum Schwinger-like process of black
hole pair creation in an external field. This process allows the
production of black holes with any mass, including masses that are
well below the Chandrasekhar limiting mass.  In order to turn a
pair of virtual black holes into a real pair one needs a
background field that provides  the energy needed to materialize
the pair, and that furnishes the force necessary to accelerate
away the black holes once they are created. In a 4-dimensional
spacetime, the pair creation process has been discussed in several
background fields, which can be: (i) an external electromagnetic
field with its Lorentz force (see \cite{Ernst}-\cite{Emparan}),
(ii) the positive cosmological constant $\Lambda$, or inflation
(see \cite{GinsPerry}-\cite{BoussoDil}), (iii) a cosmic string
with its tension (see \cite{KW}-\cite{PreskVil}), (iv) a domain
wall with its gravitational repulsive energy (see
\cite{CaldChamGibb}-\cite{Rogatko}). One can also have a
combination of the above fields, for example, a process involving
cosmic string breaking in a background magnetic field
\cite{Empar-string}, or a scenario in which a cosmic string breaks
in a cosmological background \cite{PlebDem,OscLem-PCdS}.

To study the pair creation process one needs solutions that
describe appropriately the evolution of the black hole pair after
its creation, i.e, one needs solutions that represent a pair of
black holes accelerated by the background field responsible for
the pair production. These solutions include the C-metric
\cite{KW} and the Ernst solution \cite{Ernst}, that describe a
pair accelerated by a string and by an electromagnetic field,
respectively. Unfortunately, the higher dimensional C-metric and
Ernst solutions have not been found yet. Therefore, pair creation
of higher dimensional black holes in a cosmic string background or
in an electromagnetic background cannot be discussed analytically.
However, the cosmological expansion provided by a de Sitter (dS)
background spacetime can also furnish the energy necessary for the
pair creation process. In this case, the solution that describes
the black hole pair after its creation is well known, it is the
Reissner-Nordstr\"om$-$de Sitter solution in higher dimensions
\cite{tangherlini}, and can be used to study the pair creation
process of higher dimensional black holes in a dS background. This
task will be carried in this paper. In $D=4$ ($D$ is the dimension
of the spacetime), the pair creation rates have been previously
computed in \cite{MannRoss}.

Pair creation of black holes is also a good source of
gravitational radiation emission. In an higher dimensional flat
background, an estimate for the amount of gravitational radiation
released during the pair creation period has been given in
\cite{VitOscLem}. After the pair creation, the black hole pair
accelerates away and, consequently, the black holes continue to
release energy. In a 4-dimensional dS background the gravitational
radiation emitted by uniformly accelerated black holes has been
computed in \cite{BicKrtKrtPod}.

The plan of this paper is as follows. In Sec.
 \ref{sec:Solutions D-dim} we briefly review some properties of the
 higher dimensional dS black holes that are used in the paper.
In Sec. \ref{sec:higher dimensional instantons} we construct the
instantons that describe the pair creation process. In Sec.
\ref{sec:Calc-I PC D-dim}, we explicitly evaluate the pair
creation rate of black holes. In Sec. \ref{sec:Conc} concluding
remarks are presented. Throughout this paper we use units in which
the $D$-dimensional Newton's constant is equal to one, and $c=1$.

\section{\label{sec:Solutions D-dim}Higher dimensional
\lowercase{d}S black holes}

In an asymptotically de Sitter background (positive cosmological
constant, $\Lambda> 0$), the most general static higher
dimensional black hole solution with spherical topology was found
by Tangherlini \cite{tangherlini}. The gravitational field is
\begin{equation}
ds^{2}=-f(r)dt^{2}+f(r)^{-1}dr^{2}+r^{2}\,d\Omega_{D-2}^2
 \label{RN:metric D-dim}
\end{equation}
where $D$ is the dimension of the spacetime, $d\Omega_{D-2}^2$ is
the line element on a unit $(D-2)$-sphere,
$d\Omega^2_{D-2}=d\theta_1^2+ \cdots
+\prod_{i=1}^{D-3}\sin^2\theta_i\,d\theta_{D-2}^2$, and the
function $f(r)$ is given by
\begin{equation}
f(r)=1-\frac{\Lambda}{3}r^2-\frac{M}{r^{D-3}}+\frac{Q^2}{r^{2(D-3)}}\:,
 \label{RN:f cosmolog D-dim}
\end{equation}
The mass parameter $M$ and the charge parameter $Q$ are related to
the ADM mass, $M_{\rm ADM}$, and ADM electric charge, $Q_{\rm
ADM}$, of the solution by \cite{myersperry}
\begin{eqnarray}
& &M_{\rm ADM}=\frac{(D-2)\Omega_{D-2}}{16\pi}\,M \:,\nonumber \\
& &Q_{\rm ADM}=\sqrt{\frac{(D-3)(D-2)}{2}}\,Q \:,
 \label{ADM hairs D-dim}
\end{eqnarray}
where $\Omega_{D-2}$ is the area of a unit $(D-2)$-sphere,
\begin{equation}
\Omega_{D-2}=\frac{2\pi^{(D-1)/2}}{\Gamma[(D-1)/2]}\:.
\label{integratedsolidangle}
\end{equation}
and $\Gamma[z]$ is the gamma function. The radial electromagnetic
field produced by the electric charge $Q_{\rm ADM}$ is given by
\begin{equation}
F=-\frac{Q_{\rm ADM}}{r^{D-2}}\,dt\wedge dr\:.
 \label{RN:maxwell D-dim}
\end{equation}
These solutions have a curvature singularity at the origin, and
the black hole solutions can have at most three horizons, the
Cauchy horizon $r_-$, the event horizon $r_+$ and the cosmological
horizon $r_{\rm c}$, that satisfy $r_-\leq r_+ \leq r_{\rm c}$.

\begin{figure}[ht]
\includegraphics*[height=5cm]{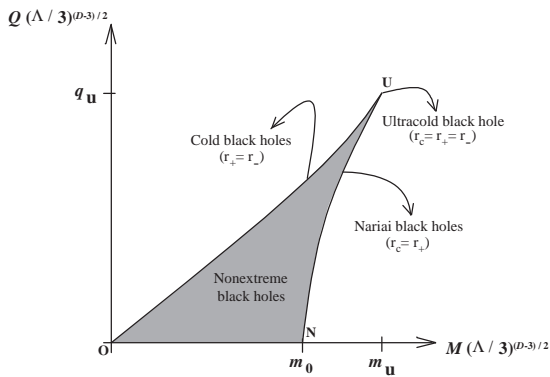}
   \caption{\label{range mq dS bh D-dim}
Range of $M$ and $Q$ for which one has a nonextreme black hole
(region interior to the closed line $ONUO$), an extreme Nariai
black hole with $r_+ = r_{\rm c}$ (line $NU$), an extreme cold
black hole with $r_- =r_+$ (line $OU$), and an extreme ultracold
black hole with $r_-=r_+ = r_{\rm c}$ (point $U$). The line $ON$
represents the nonextreme dS-Schwarzschild black hole, and point
$N$ represents the extreme Nariai Schwarzschild black hole. The
non-shaded area represents a naked singularity region. The
constants in the axes are $m_0=\frac{2}{D-1}\left (
\frac{D-3}{D-1}\right )^{(D-3)/2}$, $m_{\rm u}=\frac{4}{D-1}\left
( \frac{(D-3)^2}{(D-2)(D-1)}\right )^{(D-3)/2}$, and
 $q_{\rm u}=\frac{1}{\sqrt{D-2}}\left ( \frac{(D-3)^2}{(D-2)(D-1)}
  \right )^{(D-3)/2}$.
 }
\end{figure}

Of particular interest for us are the $D$-dimensional extreme
dS-Tangherlini black holes, for which at least two of the horizons
coincide. If we label this degenerate horizon by $\rho$, the
extreme black holes can be identified by describing the parameters
$M$ and $Q$ as a function of $\rho$ \cite{NariaiDdim}. More
concretely, the extreme black holes satisfy the relations
\cite{NariaiDdim}
\begin{eqnarray}
& &M=2\rho^{D-3} \left ( 1-\frac{D-2}{D-3}\,\frac{\Lambda}{3}
\rho^2
\right)\:, \nonumber \\
& & Q^2 =\rho^{2(D-3)} \left (
1-\frac{D-1}{D-3}\,\frac{\Lambda}{3} \rho^2 \right)\:,
 \label{mq D-dim}
 \end{eqnarray}
where the condition $Q^2\geq 0$ implies that $\rho \leq \rho_{\rm
max}$ with
\begin{eqnarray}
 \rho_{\rm max}=\sqrt{\frac{D-3}{D-1}\,\frac{3}{\Lambda}}\:.
 \label{def rho max}
\end{eqnarray}
Since a dS black hole can have at most three horizons, one can
have three distinct extreme dS black holes, namely: the cold black
hole with $r_-=r_+ \equiv \rho$, the ultracold black hole in which
the three horizons coincide, $r_-=r_+ = r_{\rm c} \equiv \rho$,
and the Nariai black hole with $r_+=r_{\rm c} \equiv \rho$ (here
we follow the nomenclature used in the analogous 4-dimensional
black holes \cite{Rom}). For the $D$-dimensional cold black hole
($r_-=r_+$), $M$ and $Q$ increase with $\rho$, and one has
\cite{NariaiDdim}
\begin{eqnarray}
 & & 0<\rho<\rho_{\rm u}\,, \qquad
 0<M<\frac{4}{D-1}\,  \rho_{\rm u}^{\, D-3}\,, \nonumber \\
 & & 0<Q<\frac{1}{\sqrt{D-2}}\, \rho_{\rm u}^{\, D-3}\:,
 \label{ColdDdim:range}
\end{eqnarray}
where we have defined
\begin{eqnarray}
 \rho_{\rm u}=\sqrt{\frac{3}{\Lambda}}\,\frac{D-3}{\sqrt{(D-2)(D-1)}}\:.
 \label{def rho ultra}
\end{eqnarray}
For the $D$-dimensional ultracold black hole
 ($r_-=r_+=r_{\rm c}$), one has \cite{NariaiDdim}
 \begin{eqnarray}
 & & \rho=\rho_{\rm u}\,, \qquad M=\frac{4}{D-1} \,
\rho_{\rm u}^{\,D-3}\,,
 \nonumber \\
 & & Q=\frac{1}{\sqrt{D-2}}\,  \rho_{\rm u}^{\,D-3}\:.
 \label{UltracoldDdim:range}
\end{eqnarray}
Finally, for the $D$-dimensional Nariai black hole
 ($r_+=r_{\rm c}$), $M$ and $Q$ decrease with $\rho$, and
one has \cite{NariaiDdim}
\begin{eqnarray}
 & & \rho_{\rm u}< \rho \leq \rho_{\rm max}\,, \quad
 \frac{2}{D-1}\,\rho_{\rm max}^{D-3}\leq M<\frac{4}{D-1} \,
\rho_{\rm u}^{\,D-3}\,,
\nonumber \\
 & &  0\leq Q<\frac{1}{\sqrt{D-2}} \, \rho_{\rm u}^{\,D-3}\:.
 \label{NariaiDdim:range}
\end{eqnarray}
The ranges of $M$ and $Q$ that represent each one of the above
extreme black holes are sketched in Fig.
 \ref{range mq dS bh D-dim}.

\section{\label{sec:higher dimensional instantons}Higher
dimensional \lowercase{d}S instantons}

In order to evaluate the black hole pair creation rate we need to
find the instantons of the theory, i.e., we must look into the
Euclidean section of the higher dimensional dS solution and choose
only those Euclidean solutions which are regular in a way that
will be explained soon. To obtain the Euclidean section of the dS
solution from the Lorentzian dS solution, (\ref{RN:metric D-dim}),
(\ref{RN:f cosmolog D-dim}) and (\ref{RN:maxwell D-dim}), we
simply introduce an imaginary time coordinate $\tau=-it$. To have
a positive definite Euclidean metric we must, in general, require
that $r$ belongs to $r_+ \leq r \leq r_{\rm c}$.

A regular instanton solution is obtained when we eliminate the
conical singularities that appear in the horizons. This is
achieved by choosing appropriately the period of $\tau$. For
example, to eliminate the conical singularity at $r=r_{\rm c}$ the
period of $\tau$ must be $\beta=2 \pi/ k_{\rm c}$, where $k_{\rm
c}$ is the surface gravity of the cosmological horizon, i.e.,
\begin{equation}
\beta=\frac{4 \pi}{|f'(r_{\rm c})|}\:,
 \label{PC D-dim:Period tau-yA PCdS}
 \end{equation}
and, in general, this period also eliminates the conical
singularity at the horizon $r_+$ if
 $|f'(r_{\rm +})|=|f'(r_{\rm c})|$.

We can construct four distinct regular instantons. Below, we will
describe each one of these four instantons, following the order:
(A) cold instanton, (B) Nariai  instanton, (C) ultracold
instanton, and (D) lukewarm  instanton. These instantons are the
higher dimensional counterparts of the 4-dimensional instantons
that have been constructed \cite{MelMos,Rom,MannRoss,BooMann} from
the Euclidean section of the dS$-$Reissner-Nordstr\"{o}m solution,
and thus we preserve the 4-dimensional nomenclature. The cold
instanton, the Nariai instanton, and the ultracold instanton are
obtained by euclideanizing solutions found in \cite{NariaiDdim}.
These four families of instantons will allow us to calculate the
pair creation rate of accelerated dS$-$Reissner-Nordstr\"{o}m
black holes in section \ref{sec:Calc-I PC D-dim}.

\subsection{\label{sec:Cold-inst PC D-dim}The higher
dimensional cold instanton}

We start with the case: $r_- =r_+$ and $r_{\rm c} \neq r_+$. This
solution is the cold instanton, and requires the presence of
charge. The gravitational field of the higher dimensional cold
instanton is given by (\ref{RN:metric D-dim}) and
 (\ref{RN:f cosmolog D-dim}), while its Maxwell field is given by
(\ref{RN:maxwell D-dim}), with the replacement $\tau=-it$.
Moreover, the degenerated horizon $\rho$, the mass parameter $M$,
and the charge parameter $Q$ satisfy relations (\ref{mq D-dim})
and (\ref{ColdDdim:range}).

In the cold instanton, the allowed range of $r$ in the Euclidean
sector is simply $r_+ < r \leq r_{\rm c}$. This occurs because
when $r_- =r_+$, the proper distance along spatial directions
between $r_+$ and $r_{\rm c}$ goes to infinity. The point
$r=r_+=\rho$ disappears from the $\tau, r$ section which is no
longer compact. Thus, in this case we have a conical singularity
only at $r_{\rm c}$, and so we obtain a regular Euclidean solution
by simply requiring that the period of $\tau$ is equal to
 (\ref{PC D-dim:Period tau-yA PCdS}). The topology of the higher dimensional
cold  instanton is ${\mathbb{R}}^2 \times S^{D-2}$ ($0 \leq \tau
\leq \beta$, $r_+ < r \leq r_{\rm c}$). The surface $r=r_+=\rho$
is then an internal infinity boundary that will have to be taken
into account in the calculation of the action of the cold
instanton (see section \ref{sec:Cold-rate PC D-dim}). The
Lorentzian sector of this cold case describes two extreme
($r_-=r_+$) dS black holes being accelerated by the cosmological
background, and the higher dimensional cold instanton describes
pair creation of these extreme black holes. To compute the pair
creation rate of cold black holes we need to know the location of
the cosmological horizon, $r_{\rm c}$. This location can be
explicitly determined for $D=4$ and $D=5$. Specifically, for $D=4$
one has $r_{\rm c}=\sqrt{3/\Lambda-2\rho^2}-\rho$, where, from
(\ref{ColdDdim:range}), one has $0<\rho<1/\sqrt{2\Lambda}$. For
$D=5$ one has $r_{\rm c}=\sqrt{3/\Lambda-2\rho^2}$, where, from
(\ref{ColdDdim:range}), one has $0<\rho<1/\sqrt{\Lambda}$. For
$D\geq 6$ finding explicitly $r_{\rm c}$ requires solving a
polynomial of degree higher than four.

\subsection{\label{sec:Nariai-inst PC D-dim}The
higher dimensional Nariai instanton}

We now turn our attention to case: $r_+ =r_{\rm c}$ and $r_{\rm -}
\neq r_+$. This solution is called Nariai instanton, and it exists
with or without charge. When the charge vanishes the only regular
Euclidean solution that can be constructed is the neutral Nariai
instanton. As we said in the beginning of this section, one
requires that $r_+ \leq r \leq r_{\rm c}$ in order to obtain a
positive definite metric. But in the Nariai case $r_+ =r_{\rm c}$,
and so it seems that we are left with no space to work with in the
Euclidean sector. However, it can be shown that the the proper
distance between $r_+$ and $r_{\rm c}$ remains finite as $r_+
\rightarrow r_{\rm c}$
\cite{GinsPerry,MannRoss,HawkRoss,NariaiDdim}.

The gravitational field of the higher dimensional Nariai instanton
 is given by \cite{NariaiDdim}
\begin{eqnarray}
d s^2 = \frac{1}{A} \left (\sin^2\chi\, d\tau^2 +d\chi^2 \right )
+ \frac{1}{B}\, d\Omega_{D-2}^2 \:.
 \label{D-dim:Nariai solution}
\end{eqnarray}
where $\chi$ runs from $0$ to $\pi$, and $A$ and $B$ are related
to $\Lambda$ and $Q$ by \cite{NariaiDdim}
\begin{eqnarray}
 & & \Lambda = \frac{3}{(D-2)(D-1)}\left [ A+(D-3)^2 B \right ]\:,
 \nonumber \\
& &  Q^2=\frac{(D-3)B-A}{(D-3)(D-2)B^{D-2}}\:.
 \label{D-dim:Nariai solution:range AB}
\end{eqnarray}
One has $B=\rho^{-2}$, where $\rho$ lies in the range defined in
(\ref{NariaiDdim:range}) \cite{NariaiDdim}. In particular, the
neutral Nariai solution satisfies $B=\rho_{\rm max}^2$, with
$\rho_{\rm max}$ defined in (\ref{def rho max}), and
$A=(D-1)\Lambda/3$. The Maxwell field (\ref{RN:maxwell D-dim}) of
the higher dimensional charged Nariai solution is
 \begin{eqnarray}
 F=i\,Q_{\rm ADM}\,\frac{B^{(D-2)/2}}{A}\,\sin \chi \,d\tau \wedge d\chi\:.
\label{D-dim:Nariai Maxwell}
\end{eqnarray}
So, if we give the parameters $\Lambda$, and $Q$ we can construct
the higher dimensional Nariai solution. The topology of the Nariai
instanton is $S^2 \times S^{D-2}$
 ($0 \leq \tau \leq \beta$, $0\leq \chi \leq \pi$).
 The Lorentzian sector of this solution is  the direct topological product of
 $dS_2 \times S^{D-2}$, i.e., of a
(1+1)-dimensional de Sitter spacetime with a  $(D-2)$-sphere of
fixed size. To each point in the sphere corresponds a $dS_2$
spacetime.  In the $D=4$ case it has been shown
\cite{GinsPerry,BoussoHawk} that the Nariai solution decays
through the quantum tunnelling process into a slightly non-extreme
dS black hole pair (for a complete review  on this subject see
\cite{Bousso60y}). We then naturally expect that an analogous
quantum instability is present in the higher dimensional Nariai
solution, as it will be shown.
Therefore, the Nariai instanton describes the creation
of a higher dimensional Nariai universe  that then decays into a
slightly non-extreme ($r_+ \sim r_{\rm c}$) pair of black holes
accelerated by the cosmological constant background.

\subsection{\label{sec:Ultracold-inst PC D-dim}The higher
dimensional ultracold instanton}

The third regular instanton is defined by (C): $r_-=r_+=r_{\rm
c}$. It is regular when condition
 (\ref{PC D-dim:Period tau-yA PCdS}) is satisfied.
This is the ultracold instanton, and can be viewed as a limiting
case of both the cold instanton and the charged Nariai instanton.

The gravitational field of the higher dimensional ultracold
instanton is given by \cite{NariaiDdim}
\begin{eqnarray}
d s^2 =\chi^2\, d\tau^2 +d\chi^2 +\rho_{\rm u}^{\,2}\,
d\Omega_{D-2}^2 \:.
 \label{D-dim:Nariai BertRob solution}
\end{eqnarray}
where $\chi$ runs from $0$ to $+\infty$, and $\rho_{\rm u}$ is
defined in (\ref{def rho ultra}). The Maxwell field of the higher
dimensional ultracold instanton  is
 \begin{eqnarray}
 F=i\,\frac{Q_{\rm ADM}}{\rho_{\rm u}^{D-2}}\,\chi \,d\tau \wedge
 d\chi\:,
\label{D-dim:Nariai BertRob Maxwell}
\end{eqnarray}
where $Q_{\rm ADM}$ is given by (\ref{ADM hairs D-dim}) and
(\ref{UltracoldDdim:range}). So, if we give  $\Lambda$ we can
construct the higher dimensional ultracold instanton. Notice that
the spacetime factor $\chi^2\, dT^2 +d\chi^2$ is just
${\mathbb{E}}^{2}$ (2-dimensional Euclidean space) in Rindler
coordinates. Therefore, under the usual coordinate transformation
$\chi=\sqrt{x^2-t^2}$ and $\tau={\rm arctanh(t/x)}$, this factor
transforms into $dt^2 +dx^2$. The topology of the ultracold
instanton is ${\mathbb{E}}^{2}\times S^{D-2}$. The Lorentzian
sector of this solution is  the direct topological product of
${\mathbb{M}}^{(1,1)}\times
S^{D-2}$, i.e., of a (1+1)-dimensional Minkowski spacetime with a
$(D-2)$-sphere of fixed size. The ultracold instanton describes the
creation of a higher dimensional Nariai$-$Bertotti-Robinson
universe \cite{NariaiDdim} that then decays into a slightly
non-extreme ($r_- \sim r_+ \sim r_{\rm c}$) pair of black holes
accelerated by the cosmological constant background.

\subsection{\label{sec:Lukewarm-inst PC D-dim}The
higher dimensional lukewarm instanton}

Finally, we have the lukewarm instanton. This instanton satisfies
$r_-\neq r_+\neq r_{\rm c}$, and
\begin{equation}
 f'(r_+)=-f'(r_{\rm c})\:.
 \label{PC D-dim:k+=kA PCdS}
 \end{equation}
In this case the surface gravities of the horizons $r_+$ and
$r_{\rm c}$ are equal ($k_+=k_{\rm c}$), and thus they have equal
Hawking temperature.  The choice
 (\ref{PC D-dim:Period tau-yA PCdS}) for the period of $\tau$ also
eliminates the conical singularity at the outer black hole
horizon, $r=r_+$. The solution is then regular in the whole
Euclidean range $r_+ \leq r \leq r_{\rm c}$.
The topology of the lukewarm instanton is $S^2 \times S^{D-2}$
($0\leq \tau \leq \beta$, $r_+ \leq r \leq r_{\rm c}$).
The Lorentzian sector
of the lukewarm solution describes two higher dimensional dS black
holes being accelerated apart by the cosmological constant, so
this instanton describes pair creation of nonextreme black holes.

The gravitational field of the higher dimensional lukewarm
instanton  is given by (\ref{RN:metric D-dim}) with the
requirement that $f(r)$ satisfies condition
 (\ref{PC D-dim:k+=kA PCdS}) and
$f(r_+)=0=f(r_{\rm c})$. To find the properties of the lukewarm
instanton we note that the function $f(r)$, given by
(\ref{RN:metric D-dim}), can also be written as
\begin{eqnarray}
& & f(r)=-\frac{\Lambda}{3}r^2 \left (  1-\frac{r_+}{r}\right )
\left
(  1-\frac{r_{\rm c}}{r}\right ) \nonumber \\
& & \hspace{1cm}\times {\biggl (}
1+\frac{a_1}{r}+\frac{a_2}{r^2}+\cdots+\frac{a_{2(D-3)}}{r^{2(D-3)}}
 {\biggr )}\:,
 \label{PC D-dim:F-luk}
 \end{eqnarray}
where $a_i$ ($i=1,\cdots,2(D-3)$) are constants that can be found
from the matching between (\ref{RN:metric D-dim}) and
 (\ref{PC D-dim:F-luk}). This matching, together with the extra condition
 (\ref{PC D-dim:k+=kA PCdS}), lead to unique relations between the
parameters $\Lambda$, $M$, $Q$  and the position of the horizons,
$r_+$ and $r_{\rm c}$. Since this procedure involves polynomials
with a high degree, we have not been able to find the general
relations between ($\Lambda$, $M$, $Q$)  and ($r_+$, $r_{\rm c}$)
for any $D$. So, we have to carry this procedure for each $D$. As
examples, we specifically discuss now the $D=4$ and the $D=5$
lukewarm instantons. For $D=4$, the above procedure yields the
relations
\begin{eqnarray}
& & \Lambda =\frac{3}{(r_{\rm c}+r_+)^2} ,
\nonumber \\
& & M=2 \frac{r_{\rm c} r_+}{r_{\rm c}+r_+}\,, \nonumber \\
& &  Q= \frac{r_{\rm c} r_+}{r_{\rm c}+r_+}\:.
 \label{PC D-dim: zeros4D}
 \end{eqnarray}
For $D=5$, the relations are
\begin{eqnarray}
& & \Lambda =3 \left ( 2r_{\rm c}+r_+ -\frac{r_{\rm c}^3(r_{\rm
c}+r_+)}{r_{\rm c}^2+r_{\rm c}r_+ + r_+^2} \right )^{-1},
\nonumber \\
& & M=\frac{r_{\rm c}^2 r_+^2 (2r_{\rm c}^2+r_{\rm c}r_+ +
2r_+^2)}
 {r_{\rm c}^4+r_{\rm c}^3r_+ +3r_{\rm c}^2 r_+^2+r_{\rm c}r_+^3 + r_+^4}\,,
\nonumber \\
& &  Q^2= \frac{r_{\rm c}^4 r_+^4}
 {r_{\rm c}^4+r_{\rm c}^3r_+ +3r_{\rm c}^2 r_+^2+r_{\rm c}r_+^3 + r_+^4}\:.
 \label{PC D-dim: zeros5D}
 \end{eqnarray}
These two examples indicate an important difference between the
lukewarm instanton in $D=4$ dimensions and in $D\geq 5$: for $D=4$
the lukewarm instanton has a ADM mass, $M_{\rm ADM}=M/2$, equal to
its ADM charge, $Q_{ADM}=Q$, while for $D\geq 5$ one has $M_{\rm
ADM}\neq Q_{\rm ADM}$.
Note also that relations (\ref{PC D-dim: zeros4D})
and (\ref{PC D-dim: zeros5D})
and their higher dimensional counterparts define implicitly
$r_{\rm c}$ and $r_+$ as a function of $\Lambda$, $M$, and $Q$.
The location of $r_{\rm c}$ and $r_+$ can be explicitly determined
for $D=4$ and $D=5$.  For $D\geq 6$ finding explicitly $r_{\rm c}$
requires solving a polynomial of degree higher than four.

\section{\label{sec:Calc-I PC D-dim}Calculation of the black hole
pair creation rates}

The pair creation rate of higher dimensional black holes in a dS
background is given, according to the no-boundary proposal of
\cite{HartleHawk}, by
\begin{eqnarray}
\Gamma =\eta \, e^{-2I_{\rm inst}+2I_{\rm dS}} \:,
 \label{PC D-dim:PC-rate}
 \end{eqnarray}
where $\eta$ is the one-loop contribution from the quantum
quadratic fluctuations in the fields that will not be considered
here (for the computation of this factor in some special cases see
\cite{GinsPerry,VolkovWipf,OneLoop}). $I_{\rm inst}$ is the
classical Euclidean action of the gravitational instanton that
mediates the pair creation of black holes, given by
\cite{Brown,HawkRoss}
 \begin{eqnarray}
I_{\rm inst}&=&-\frac{1}{16\pi}\int_{\cal M} d^Dx\sqrt{g} \left (
R-2\lambda-F^{\mu\nu}F_{\mu\nu} \right ) \nonumber \\
& & -\frac{1}{8\pi}\int_{\Sigma=\partial {\cal M}}
d^{D-1}x\sqrt{h}\, K  \nonumber \\
 & & -\frac{1}{4\pi}\int_{\Sigma=\partial {\cal M}}
d^{D-1}x\sqrt{h}\, F^{\mu\nu}n_{\mu}A_{\nu} \:,
 \label{PC D-dim:I-electric}
 \end{eqnarray}
where $\Sigma=\partial {\cal M}$ is the boundary of a compact
manifold ${\cal M}$, $g$ is the determinant of the Euclidean
metric, $h$ is the determinant of the induced metric on the
boundary $\Sigma$, $R$ is the Ricci scalar, $K$ is the trace of
 the extrinsic curvature $K_{ij}$ of the boundary,
 $F_{\mu\nu}=\partial_{\mu}A_{\nu}-\partial_{\nu}A_{\mu}$ is
the Maxwell field strength of the gauge field $A_{\nu}$, $n_{\mu}$
is the unit outward normal to $\Sigma$, and we have defined
\begin{eqnarray}
\lambda=\frac{(D-1)(D-2)\,\Lambda}{6} \:.
 \label{D-dim:def lambda}
\end{eqnarray}
The coefficient of the $\Lambda$ term was chosen in order to
insure that, for any dimension $D$, the pure dS spacetime is
described by $f(r)=1-(\Lambda/3)r^2$, as occurs with $D=4$. The
expression for the Ricci scalar in $D$-dimensions is worth of a
comment. Variation of the action (\ref{PC D-dim:I-electric})
yields the equation for the gravitational field,
$R_{\mu\nu}-\frac{1}{2}R\,g_{\mu\nu}+\lambda
g_{\mu\nu}=8\pi\,T_{\mu\nu}$,  where $R_{\mu\nu}$ is the Ricci
tensor and $T_{\mu\nu}$ is the electromagnetic energy-momentum
tensor, $T_{\mu\nu}=\frac{1}{4\pi}\left ( g^{\alpha \beta}
F_{\alpha \mu} F_{\beta
\nu}-\frac{1}{4}g_{\mu\nu}F_{\alpha\beta}F^{\alpha\beta} \right
)$. The contraction of the Einstein equation with $g^{\mu\nu}$
yields the Ricci tensor
\begin{eqnarray}
R=\frac{D(D-1)}{3}\,\Lambda-\frac{16\pi}{D-2}\,T \:,
 \label{D-dim:Ricci Tensor}
\end{eqnarray}
where $T$ is the trace of $T_{\mu\nu}$.
 We note that in a general $D$-dimensional background the electromagnetic
energy-momentum tensor is not traceless. Indeed, the contraction
of $T_{\mu\nu}$ with $g^{\mu\nu}$ yields
\begin{eqnarray}
T=-\frac{D-4}{4\pi}\,F_{\mu\nu}F^{\mu\nu} \:,
 \label{D-dim:trace T}
\end{eqnarray}
which vanishes only for $D=4$.

In (\ref{PC D-dim:PC-rate}), $I_{\rm dS}$ is the Euclidean action
of the $S^{D}$ gravitational instanton that mediates the
nucleation of a dS space from nothing. Use of
 (\ref{RN:f cosmolog D-dim}), (\ref{PC D-dim:Period tau-yA PCdS})
and (\ref{D-dim:Ricci Tensor}), with $M=0$ and $Q=0$, yields for
$I_{\rm dS}$ the value
\begin{eqnarray}
 I_{\rm dS}&=& -\frac{1}{16\pi}\int d^D x \sqrt{g} \left (
R-2\lambda \right ) \nonumber \\
&=& -\frac{1}{4}\left( \frac{3}{\Lambda}\right)^{(D-2)/2}\!\!
\frac{\pi^{(D-1)/2}}{\Gamma[(D-1)/2]} \:.
 \label{PC D-dim:I dS pure}
\end{eqnarray}

\subsection{\label{sec:Cold-rate PC D-dim}The higher dimensional
cold pair creation rate}

The Maxwell field of the higher dimensional cold is $F=-i
\frac{Q_{\rm ADM}}{r^{D-2}}\, d\tau \wedge dr$. With this
information we are able to compute all the terms of the Euclidean
action (\ref{PC D-dim:I-electric}). We start with
\begin{widetext}
  \begin{eqnarray}
\hspace{-1.0 cm} -\frac{1}{16\pi}\int_{\cal{M}} d^Dx\sqrt{g} \left
( R-2\lambda \right ) &=& \left(
-\frac{D-1}{24\pi}\,\Lambda+\frac{(D-4)(D-3)}{16 \pi}\,Q^2 \right
) \int d\Omega_{D-2}
 \int_0^{\beta/2} \!\!\!\!d\tau
\int_{\rho}^{r_{\rm c}} \!\!\!\!dr\: r^{D-2}
 \nonumber \\
&=& \frac{\pi^{(D-3)/2}}{\Gamma[(D-1)/2]}\,\frac{\beta}{8}
 \left [ -\frac{\Lambda}{3}
 \left ( r_{\rm c}^{\,D-1}-\rho^{\,D-1} \right )
+\frac{(D-4)\,Q^2}{2}
 \left (\rho^{\,-(D-3)}-r_{\rm c}^{\,-(D-3)} \right ) \right ]
\:,
 \label{PC D-dim:I1-cold}
   \end{eqnarray}
 \end{widetext}
where  $\int d\Omega_{D-2}=\Omega_{D-2}$ is defined in
(\ref{integratedsolidangle}). The Maxwell term in the action
yields
\begin{eqnarray}
& &\hspace{-0.4 cm} \frac{1}{16\pi}\int_{\cal{M}} d^D x\sqrt{g}
\:F^2 \nonumber \\
& & =-\frac{(D-2)\,Q^2\,\beta}{16}
\frac{\pi^{(D-3)/2}}{\Gamma[(D-1)/2]}
 \left [\rho^{\,-(D-3)}-r_{\rm c}^{\,-(D-3)} \right ]\!, \nonumber \\
& &
 \label{PC D-dim:I2-cold}
 \end{eqnarray}
and $\int_{\Sigma} d^{D-1}x\sqrt{h}\, K=0$. In order to compute
the extra Maxwell boundary term in (\ref{PC D-dim:I-electric}) we
have to find a vector potential, $A_{\nu}$, that is regular
everywhere including at the horizons. An appropriate choice in the
cold case is $A_r=- i\,\frac{Q_{\rm ADM}}{r^{D-2}}\,\tau$. The
integral over $\Sigma$ consists of an integration between $\rho$
and $r_{\rm c}$ along the $\tau=0$ surface and back along
$\tau=\beta/2$, and of an integration between $\tau=0$ and
$\tau=\beta/2$ along the $r=r_{\rm c}$ surface and back along the
$r=\rho$ surface. The normal to $\Sigma_{\tau}$ is
 $n_{\mu}=\left (\sqrt{f(r)},0,\cdots,0\right )$, and the normal
to $\Sigma_{h}$ is
$n_{\mu}=\left (0,\sqrt{f(r)},0,\cdots,0\right )$. Thus the
non-vanishing contribution comes only from the integration along
the $\tau=\beta/2$ surface. The Maxwell boundary term in
 (\ref{PC D-dim:I-electric}) is then
 \begin{eqnarray}
\hspace{-0.5cm}-\frac{1}{4\pi}\int_{\Sigma_{\tau=\beta/2}}
\!\!\!\!\!\!\!\!\!\!\!\!
d^{D-1}x\sqrt{h}\, F^{\tau r}n_{\tau}A_{r} =
-\frac{1}{8\pi}\int_{\cal{M}}\!\!\! d^D x\sqrt{g} \:F^2 .
 \label{PC D-dim:I-electric-cold}
 \end{eqnarray}
Adding all these terms yields the action
 (\ref{PC D-dim:I-electric}) of the higher dimensional cold
instanton (onwards the subscript ``${\rm c}$" means cold)
\begin{eqnarray}
 I_{\rm c}= -\frac{r_{\rm c}^{\,D-2}}{4}\,
\frac{\pi^{(D-1)/2}}{\Gamma[(D-1)/2]}
 \:, \label{PC D-dim:I-total-cold}
 \end{eqnarray}
which, for $D=4$, reduces to the result of \cite{MannRoss}. The
allowed interval of $\rho$ is defined in (\ref{ColdDdim:range}).
As $\rho$ varies from $\rho=\rho_{\rm u}$, defined in
 (\ref{def rho ultra}), to $\rho=0$, the cold action
(\ref{PC D-dim:I-total-cold})
varies according to
\begin{eqnarray}
 -\frac{ \rho_{\rm u}^{\, D-2} }{4}
\frac{\pi^{(D-1)/2}}{\Gamma[(D-1)/2]} < I_{\rm c} < I_{\rm dS} \:,
  \label{PC D-dim:rangeI-cold}
 \end{eqnarray}
where the lower limit of this relation is the ultracold action, as
we shall see in (\ref{PC D-dim:I-total-ultracold}), and $I_{\rm
dS}$ is defined in (\ref{PC D-dim:I dS pure}).

The pair creation rate of extreme cold black holes is given by
(\ref{PC D-dim:PC-rate}),
\begin{eqnarray}
\Gamma_{\rm c}=\eta\,e^{-2I_{\rm c}+2I_{\rm dS}}\,,
 \label{PC rate D-dim:cold}
 \end{eqnarray}
where $\eta$ is the one-loop contribution not computed here.

\centerline{}
\centerline{}
\subsection{\label{sec:Nariai-rate PC D-dim}The higher
dimensional Nariai pair creation rate}

 The first term of the Euclidean action
(\ref{PC D-dim:I-electric}) gives in the Nariai case
\begin{widetext}
  \begin{eqnarray}
 -\frac{1}{16\pi}\int_{\cal{M}} d^Dx\sqrt{g} \left ( R-2\lambda \right )
&=& \left (-\frac{\Lambda\,(D-1)}{24\pi}+\frac{(D-4)(D-3)}{16\pi}
 \,Q^2 B^{D-2} \right ) \int d\Omega_{D-2}
 \int_0^{2\pi /2} \!\!\!\!d\tau
\int_{0}^{\pi} \!\!\!\!d\chi\:\frac{\sin \chi}{A\,B^{(D-2)/2}} \nonumber \\
&=& \frac{\pi^{(D-1)/2}}{\Gamma[(D-1)/2]}
 \left [ -\frac{\Lambda\,(D-1)}{6}\, \frac{1}{A\,B^{(D-2)/2}}
 + \frac{Q^2 \,(D-4)(D-3)}{4}\, \frac{B^{(D-2)/2}}{A}\right ]\:.
 \label{PC D-dim:I1-Nariai}
 \end{eqnarray}
 \end{widetext}
The Maxwell term in the action yields
\begin{eqnarray}
 & & \hspace{-0.8 cm} \frac{1}{16\pi}\int_{\cal{M}} d^D x\sqrt{g}
\:F^2 \nonumber \\
& &\hspace{-0.6 cm} =-\frac{(D-2)(D-3)\,Q^2}{4}\,
\frac{B^{(D-2)/2}}{A}\, \frac{\pi^{(D-1)/2}}{\Gamma[(D-1)/2]},
 \label{PC D-dim:I2-Nariai}
 \end{eqnarray}
and $\int_{\Sigma} d^{D-1}x\sqrt{h}\, K=0$. In order to compute
the extra Maxwell boundary term in (\ref{PC D-dim:I-electric}) we
have to find a vector potential, $A_{\nu}$, that is regular
everywhere including at the horizons. An appropriate choice in the
Nariai case is $A_{\chi}= i\,Q_{\rm ADM}\,\frac{B^{(D-2)/2}}{A}
\sin \chi \:\tau$. The integral over $\Sigma$ consists of an
integration between $\chi=0$ and $\chi=\pi$ along the $\tau=0$
surface and back along $\tau=\pi$, and of an integration between
$\tau=0$ and $\tau=\pi$ along the $\chi=0$ surface, and back along
the $\chi=\pi$ surface. The non-vanishing contribution to the
Maxwell boundary term in (\ref{PC D-dim:I-electric}),
$-\frac{1}{4\pi}\int_{\Sigma} d^3x\sqrt{h}\, F^{\mu\nu}n_{\mu}
A_{\nu}$, comes only from the integration along the $\tau=\pi$
surface and is given by
 \begin{eqnarray}
\hspace{-0.5cm}-\frac{1}{4\pi}\int_{\Sigma_{\tau=\pi}}\!\!\!\!\!\!\!\!
d^{D-1}x\sqrt{h}\, F^{\tau \chi}n_{\tau}A_{\chi}=
-\frac{1}{8\pi}\int_{\cal{M}} \!\!\! d^D x\sqrt{g} \:F^2 .
 \label{PC D-dim:I-electric-Nariai}
 \end{eqnarray}
Adding all these terms yields the action
 (\ref{PC D-dim:I-electric}) of the higher dimensional Nariai instanton
(onwards the subscript ``${\rm N}$" means Nariai)
\begin{eqnarray}
 I_{\rm N}= -\frac{1}{2\,B^{(D-2)/2}}
\frac{\pi^{(D-1)/2}}{\Gamma[(D-1)/2]} \:,
  \label{PC D-dim:I-total-Nariai}
 \end{eqnarray}
which, for $D=4$, reduces to the result of
\cite{HawkRoss,MannRoss}. One has $B=\rho^{-2}$, where $\rho$ lies
in the range defined in (\ref{NariaiDdim:range}). Thus, the Nariai
action
 (\ref{PC D-dim:I-total-Nariai}) lies in the range
\begin{eqnarray}
-\frac{\pi^{(D-1)/2}\, \rho_{\rm max}^{D-2}}{2\, \Gamma[(D-1)/2]}
  \leq I_{\rm N}
 < - \frac{\pi^{(D-1)/2}\, \rho_{\rm u}^{D-2}}{2\, \Gamma[(D-1)/2]}
  \:,
  \label{PC D-dim:rangeI-Nariai}
 \end{eqnarray}
where the quantities $\rho_{\rm max}$ and $\rho_{\rm u}$ are
defined, respectively, in (\ref{def rho max}) and (\ref{def rho
ultra}). The equality holds in the neutral Nariai case ($Q=0$),
and this case has been previously discussed in \cite{paul}, while
the upper limit of (\ref{PC D-dim:rangeI-Nariai}) is twice the value
of the ultracold action, which will be defined in
(\ref{PC D-dim:I-total-ultracold}).

The pair creation rate of extreme Nariai black holes is given by
(\ref{PC D-dim:PC-rate}),
\begin{eqnarray}
\Gamma_{\rm N}=\eta\,e^{-2I_{\rm N}+2I_{\rm dS}}\,,
 \label{PC rate D-dim:Nariai}
 \end{eqnarray}
where $I_{\rm dS}$ is given by (\ref{PC D-dim:I dS pure}), and
$\eta$ is the one-loop contribution not computed here. The process
studied in this subsection describes the nucleation of a higher
dimensional Nariai universe that is unstable
\cite{GinsPerry,BoussoHawk,Bousso60y} and decays through the pair
creation of extreme Nariai black holes.

\subsection{\label{sec:Ultracold-rate PC D-dim}The higher
dimensional ultracold pair creation rate}

 The boundary $\Sigma=\partial \cal{M}$ that
appears in (\ref{PC D-dim:I-electric}) consists of an initial
spatial surface at $\tau=0$ plus a final spatial surface at
$\tau=\pi$. We label these two 3-surfaces by $\Sigma_{\tau}$. Each
one of these two spatial $(D-1)$-surfaces is bounded by a
($D-2$)-surface at the Rindler horizon $\chi=0$ and by a
($D-2$)-surface at the internal infinity $\chi=\infty$. The two
surfaces $\Sigma_{\tau}$ are connected by a timelike
($D-1$)-surface, $\Sigma_{h}$, that intersects $\Sigma_{\tau}$ at
the frontier $\chi=0$ and by a timelike ($D-1$)-surface,
$\Sigma^{\rm int}_{\infty}$, that intersects $\Sigma_{\tau}$ at
the internal infinity boundary $\chi=\infty$. Thus
$\Sigma=\Sigma_{\tau}+\Sigma_{h}+\Sigma^{\rm int}_{\infty}$, and
the region $\cal{M}$ within it is compact. The first term of the
Euclidean action (\ref{PC D-dim:I-electric}) yields
\begin{widetext}
 \begin{eqnarray}
 -\frac{1}{16\pi}\int_{\cal{M}} d^Dx\sqrt{g} \left ( R-2\lambda \right )
& &= \frac{1}{16\pi}\left [ \Lambda \,\frac{2(D-1)}{3}
 \, \rho_{\rm u}^{\, D-2} + Q^2\,(D-4)(D-3)\frac{1}{\rho_{\rm u}^{\, D-2} }
\right ] \int d\Omega_{D-2}
 \int_0^{2\pi /2} \!\!\!\!d\tau
\int_{0}^{\chi_0\rightarrow \infty} \!\!\!\!d\chi \, \chi \nonumber \\
 & &=  \left [ -\frac{D-1}{24}\, \Lambda\, \rho_{\rm u}^{\, D-2}
 + \frac{(D-4)(D-3)}{16}\,\frac{Q^2}{\rho_{\rm u}^{\, D-2}}\right ]\,
\frac{\pi^{(D-1)/2}}{\Gamma[(D-1)/2]}\, \chi_0^{\,2}
 {\biggl |}_{\chi_0\rightarrow \infty} \:.
 \label{PC D-dim:I1-ultracold}
 \end{eqnarray}
\end{widetext}
 The Maxwell term in the action
yields
\begin{eqnarray}
 & & \hspace{-0.8 cm} \frac{1}{16\pi}\int_{\cal{M}} d^D x\sqrt{g} \:F^2
\nonumber \\
 & & =-\frac{(D-3)(D-2)\,Q^2}{16\, \rho_{\rm u}^{\, D-2}}
\frac{\pi^{(D-1)/2}}{\Gamma[(D-1)/2]} \chi_0^{\,2}
 {\biggl |}_{\chi_0\rightarrow \infty} \!\!\!\!\!\!.
 \label{PC D-dim:I2-ultracold}
 \end{eqnarray}

Now, contrary to the other instantons, the ultracold instanton has
a non-vanishing extrinsic curvature boundary term,
$-\frac{1}{16\pi}\int_{\Sigma} d^{D-1}x\sqrt{h}\, K \neq 0$, due
to the internal infinity boundary ($\Sigma^{\rm int}_{\infty}$ at
$\chi=\infty$) contribution. The extrinsic curvature to
$\Sigma^{\rm int}_{\infty}$ is
$K_{\mu\nu}=h_{\mu}^{\:\:\:\alpha}\nabla_{\alpha}n_{\nu}$, where
 $n_{\nu}=(0,1,0,\cdots,0)$ is the unit outward normal to
$\Sigma^{\rm int}_{\infty}$,
$h_{\mu}^{\:\:\:\alpha}=g_{\mu}^{\:\:\:\alpha}-n_{\mu}n^{\alpha}
 =(1,0,1,\cdots,1)$ is the projection tensor onto $\Sigma^{\rm int}_{\infty}$,
 and $\nabla_{\alpha}$ represents the covariant derivative with respect
 to $g_{\mu\nu}$. Thus the trace of the extrinsic
curvature to $\Sigma^{\rm int}_{\infty}$ is
$K=g^{\mu\nu}K_{\mu\nu}=\frac{1}{\chi}$, and
\begin{eqnarray}
 -\frac{1}{8\pi}\int_{\Sigma} d^{D-1}x\sqrt{h}\, K=
 -\frac{ \rho_{\rm u}^{\, D-2} }{4}
\frac{\pi^{(D-1)/2}}{\Gamma[(D-1)/2]} \:.
 \label{PC D-dim:I3-ultracold}
 \end{eqnarray}
In the ultracold case the vector potential $A_{\nu}$, that is
regular everywhere including at the horizon, needed to compute the
extra Maxwell boundary term in (\ref{PC D-dim:I-electric}) is
$A_{\tau}=- i\,\frac{Q_{\rm ADM}}{\rho^{D-2}}\,\chi^2/2$. The
integral over $\Sigma$ consists of an integration between $\chi=0$
and $\chi=\infty$ along the $\tau=0$ surface and back along
$\tau=\pi$, and of an integration between $\tau=0$ and $\tau=\pi$
along the $\chi=0$ surface, and back along the internal infinity
 surface $\chi=\infty$. The non-vanishing contribution to the Maxwell
boundary term in (\ref{PC D-dim:I-electric}) comes only from the
integration along the internal infinity boundary $\Sigma^{\rm
int}_{\infty}$, and is given by
 \begin{eqnarray}
\hspace{-0.5cm} -\frac{1}{4\pi}\int_{\Sigma^{\rm int}_{\infty}}
\!\!\!\!\!\! d^{D-1}x\sqrt{h}\, F^{\chi\tau}n_{\chi}A_{\tau}=
-\frac{1}{8\pi}\int_{\cal{M}} \!\!\! d^D x\sqrt{g} \:F^2 .
 \label{PC D-dim:I-electric-ultracold}
 \end{eqnarray}
Due to the fact that $\chi_0\rightarrow \infty$ it might seem that
the contribution from (\ref{PC D-dim:I1-ultracold}), (\ref{PC
D-dim:I2-ultracold}) and (\ref{PC D-dim:I-electric-ultracold})
diverges. This is not however the case since these three terms
cancel each other. The only contribution to the action
 (\ref{PC D-dim:I-electric}) of the higher dimensional ultracold instanton
(onwards the subscript ``${\rm u}$" means ultracold) comes from
(\ref{PC D-dim:I3-ultracold}) yielding
\begin{eqnarray}
I_{\rm u}=
  -\frac{ \rho_{\rm u}^{\, D-2} }{4}
\frac{\pi^{(D-1)/2}}{\Gamma[(D-1)/2]}\:,
  \label{PC D-dim:I-total-ultracold}
 \end{eqnarray}
which, for $D=4$, reduces to the result of \cite{MannRoss}. The
ultracold action coincides with the minimum value of the cold
action range (\ref{PC D-dim:rangeI-cold}), and is equal to one
half the maximum value of the Nariai action range
 (\ref{PC D-dim:rangeI-Nariai}).

The pair creation rate of extreme ultracold black holes is given
by (\ref{PC D-dim:PC-rate}),
\begin{eqnarray}
\Gamma_{\rm u}=\eta\,e^{-2I_{\rm u}+2I_{\rm dS}}\,,
 \label{PC rate D-dim:ultracold}
 \end{eqnarray}
where $I_{\rm dS}$ is given by (\ref{PC D-dim:I dS pure}), and
$\eta$ is the one-loop contribution not computed here. The process
studied in this subsection describes the nucleation of a higher
dimensional Nariai$-$Bertotti-Robinson universe that is unstable,
and decays through the pair creation of extreme ultracold black
holes.

\subsection{\label{sec:Lukewarm-rate PC D-dim}The higher dimensional
lukewarm  pair creation rate}

The evaluation of the Euclidean action of the higher dimensional
lukewarm instanton follows as in the cold case as long as we
replace $\rho$ by $r_+$. Therefore, the action
 (\ref{PC D-dim:I-electric}) of the higher dimensional lukewarm
instanton (onwards the subscript ``$\ell$" means lukewarm) is given
by\footnote{We thank Alexander Vilenkin for pointing out a typo in
(\ref{PC D-dim:I-total-luk}) in a previous version of the paper.}
\begin{eqnarray}
 I_{\rm \ell}&=&
 \frac{\pi^{(D-3)/2}}{\Gamma[(D-1)/2]}\,\frac{\beta}{8}
 {\biggl [}
 -\frac{\Lambda}{3} \left ( r_{\rm c}^{\,D-1}-r_+^{\,D-1} \right )
  \nonumber \\
 & & +(D-3)Q^2
 \left (r_+^{\,-(D-3)}-r_{\rm c}^{\,-(D-3)} \right )
 {\biggr ]}\:. \label{PC D-dim:I-total-luk}
 \end{eqnarray}
and the pair creation rate of nonextreme lukewarm black holes is
given by (\ref{PC D-dim:PC-rate}).

\subsection{\label{sec:Sub-Maximal-rate D-dim}Pair creation rate of higher
dimensional nonextreme sub-maximal black holes}

The cold, Nariai, ultracold and lukewarm instantons are saddle
point solutions free of conical singularities both in the $r_+$
and $r_{\rm c}$ horizons. The corresponding black holes may then
nucleate in the dS background, and we have computed their pair
creation rates in the last four subsections. However, these
particular black holes are not the only ones that can be pair
created. Indeed, it has been shown in
\cite{WuSubMaxBoussoHawkSubMax} that Euclidean solutions with
conical singularities may also be used as saddle points for the
pair creation process. In this way, pair creation of nonextreme
sub-maximal black holes is allowed (by this nomenclature we mean
all the nonextreme black holes other than the lukewarm ones that
are in the region interior to the close line $ONUO$ in Fig.
\ref{range mq dS bh D-dim}), and their pair creation rate may be
computed. In order to calculate this rate, the action is given by
(\ref{PC D-dim:I-electric}) and, in addition, it has now an extra
contribution from the conical singularity (c.s.) that is present
in one of the horizons ($r_+$, say) given by
\cite{ReggeGibbonsPerryAconSing,GinsPerry}
\begin{eqnarray}
\frac{1}{16\pi}\int_{\cal{M}} d^Dx\sqrt{g}
 \:\left (
R-2\lambda \right ){\biggl |}_{{\rm c.s.}\:{\rm at}\:r_+}
 \!\!\!= \frac{ {\cal A}_+\:\delta}{4\,D\,\pi}\:,
 \label{I conical sing D-dim}
 \end{eqnarray}
where ${\cal A}_+=\frac{2\pi^{(D-1)/2}}{\Gamma[(D-1)/2]}
\,r_+^{D-2}$ is the area of the ($D-2$)-sphere spanned by the
conical singularity, and
\begin{eqnarray}
\delta=2\pi \left ( 1-\frac{\beta_{\rm c}}{\beta_+}\right )
 \label{delta concical sing PCdS D-dim}
 \end{eqnarray}
is the deficit angle associated to the conical singularity at the
horizon $r_+$, with $\beta_{\rm c}=4 \pi / |f'(r_{\rm c})|$ and
$\beta_+=4 \pi / |f'(r_+)|$ being the periods of $\tau$ that avoid
a conical singularity in the horizons $r_{\rm c}$ and $r_+$,
respectively. The contribution from (\ref{PC D-dim:I-electric})
follows straightforwardly in a similar way as the one shown in
subsection \ref{sec:Lukewarm-rate PC D-dim} with the period of
$\tau$, $\beta_{\rm c}$,  chosen in order to avoid the conical
singularity at the cosmological horizon, $r=r_{\rm c}$.

\section{\label{sec:Heuristic}Heuristic derivation of the
pair creation rates}

The physical interpretation of our exact results can be clarified
with a heuristic derivation of the nucleation rates. An estimate
for the nucleation probability is given by the Boltzmann factor,
$\Gamma \sim e^{-E_0/W_{\rm ext}}$, where $E_0$ is the energy of
the system that nucleates and $W_{\rm ext}=F \ell$ is the work
done by the external force $F$, that provides the energy for the
nucleation, through the typical distance $\ell$ separating the
created pair. We can then show that the creation probability for a
black hole pair in a dS background is given by $\Gamma \sim
e^{-M/\sqrt{\Lambda}}$, in agreement with the exact results.
Indeed, one has $E_0 \sim 2M$, where $M$ is the rest energy of the
black hole, and $W_{\rm ext}\sim \sqrt{\Lambda}$ is the work
provided by the cosmological background. To derive $W_{\rm
ext}\sim \sqrt{\Lambda}$ one can argue as follows. In the dS case,
the Newtonian potential is $\Phi=\Lambda r^2/3$ and its derivative
yields the force per unit mass or the acceleration $a$, $\Lambda
r$.  This acceleration should be evaluated at the characteristic
dS radius, $r=1/\sqrt{\Lambda}$, yielding $a=\sqrt\Lambda$.  The
force can then be written as $F= {\rm mass}\times{\rm
acceleration}\sim \sqrt{\Lambda}\sqrt{\Lambda}$, where the
characteristic mass of the system is $\sqrt{\Lambda}$. Thus, the
characteristic work is $W_{\rm ext}={\rm force}\times{\rm
distance}\sim \Lambda\,(1/\sqrt \Lambda)\sim \sqrt{\Lambda}$,
where the characteristic distance that separates the pair at the
creation moment is $\sqrt\Lambda$ (This value follows from the
fact that the dS spacetime can be represented as a hyperboloid in
a Minkowski embedding spacetime, and the origin of the dS
spacetime describes the hyperbolic trajectory, $X^2-T^2=\Lambda$,
in the embedding space). So, from the Boltzmann factor we indeed
expect that the creation rate of a black hole pair in a dS
background is given by $\Gamma \sim e^{-M/\sqrt{\Lambda}}$. This
expression is in agreement with our results since from (\ref{mq
D-dim}), one has $M \sim \rho^{D-3} \sim \Lambda^{-(D-3)/2}$, and
thus $\Gamma \sim e^{-M/\sqrt{\Lambda}}\sim
e^{-\Lambda^{-(D-2)/2}}$.

\section{\label{sec:Conc}Discussion of the results}

We have studied in detail the quantum process in which a pair of
black holes is created in a higher dimensional de Sitter (dS)
background, a process that in $D=4$ was previously discussed in
\cite{MannRoss}. The energy to materialize and accelerate the pair
comes from the positive cosmological constant. The dS space is the
only background in which we can discuss analytically the pair
creation process of higher dimensional black holes, since the
C-metric and the Ernst solutions, that describe respectively a
pair accelerated by a string and by an electromagnetic field, are
not know yet in a higher dimensional spacetime.

As occurs for $D=4$, the pair creation of higher
dimensional black holes is always suppressed relative to the dS
space, i.e., the argument of the exponential function that defines
the pair creation rate (\ref{PC D-dim:PC-rate}) is always
negative. To compare the evolution of the pair creation rate as
the dimension of the spacetime increases, we note that this rate
goes as $\Gamma \sim e^{-\Lambda^{-(D-2)/2}}$. Thus, for a fixed
value of $\Lambda$, the rate increases when $D$ grows, i.e., pair
creation of black holes in the dS background becomes less
suppressed when the dimension of the spacetime increases. This
behavior increases the interest of this kind of black hole
creation process in a higher dimensional spacetime.

In previous works on black hole pair creation in general
background fields it has been well established that the pair
creation rate is proportional to the exponential of the
gravitational entropy $S$ of the system, $\Gamma \propto e^S$,
with the entropy being given by one quarter of the the total area
$\cal{A}$ of all the horizons present in the instanton,
$S={\cal{A}}/4$. It is straightforward to verify that these
relations also hold for the higher dimensional dS instantons \cite{Wu}.
Indeed, in the cold case, the instanton has a single horizon,  the
cosmological horizon at $r=r_{\rm c}$, in its Euclidean section,
since $r=r_+$ is an internal infinity. So, the total area of the
cold instanton is ${\cal A}_{\rm c}=\Omega_{D-2}\,r_{\rm
c}^{D-2}$. Thus, $S_{\rm c}=-2I_{\rm c}={\cal{A}_{\rm c}}/4$,
where $I_{\rm c}$ is given by (\ref{PC D-dim:I-total-cold}). In
the Nariai case, the instanton has two horizons in its Euclidean
section, namely the cosmological horizon $r=r_{\rm c}$ and the
black hole horizon $r_+$, both at $r=\rho=B^{-1/2}$, and thus they
have the same area. So, the total area of the Nariai instanton is
${\cal A}_{\rm N}=2\Omega_{D-2}\,B^{-(D-2)/2}$. Again, one has
$S_{\rm N}=-2I_{\rm N}={\cal{A}_{\rm N}}/4$, where $I_{\rm N}$ is
given by (\ref{PC D-dim:I-total-Nariai}). In the ultracold  case,
the instanton has a single horizon in its Euclidean section, the
Rindler horizon at $\chi=0$, since
 $\chi=\infty$ is an internal infinity. The total area of the
ultracold instanton is then ${\cal A}_{\rm
u}=\Omega_{D-2}\,\rho_{\rm u}^{D-2}$. Thus, $S_{\rm u}=-2I_{\rm
u}={\cal{A}_{\rm u}}/4$, where $I_{\rm u}$ is given by (\ref{PC
D-dim:I-total-ultracold}).

The ultracold  instanton is a limiting case of both the charged
Nariai instanton and the cold instanton (see Fig.
 \ref{range mq dS bh D-dim}). Then, as expected, the action of the
cold instanton gives, in this limit, the action of the ultracold
instanton (see (\ref{PC D-dim:rangeI-cold})). However, the
ultracold frontier of the Nariai action is given by two times the
ultracold action (see (\ref{PC D-dim:rangeI-Nariai})). The reason
for this behavior is clear. Indeed, in the ultracold case and in
the cold case, the respective instantons have a single horizon
(the other possible horizon turns out to be a internal infinity).
This horizon gives the only contribution to the total area,
${\cal{A}}$, and therefore to the pair creation rate. In the
Nariai case, the instanton has two horizons with the same area,
and thus the ultracold limit of the Nariai action is twice
the value of the true ultracold action.


\begin{acknowledgments}

This work was partially funded by Funda\c c\~ao para a Ci\^encia e
Tecnologia (FCT) through project CERN/FIS/43797/2001.
OJCD acknowledges financial support from FCT
through grant SFRH/BPD/2003. JPSL thanks Observat\'orio Nacional
do Rio de Janeiro for hospitality.

\end{acknowledgments}



\end{document}